\documentclass{iopart}

\usepackage{iopams}
\usepackage{graphicx,subfigure}
\usepackage{cite}

\newcommand{\avg}[1]{\left\langle{#1}\right\rangle}
\newcommand{\ovl}[1]{\overline{#1}}
\newcommand{\ii}{{\rm i}}

\newcommand{\erf}{{\rm erf}}

\renewcommand{\l}{\left}
\renewcommand{\r}{\right}

\begin{document}

\title{Statistical mechanics analysis of the equilibria of linear
economies}

\author{A De Martino\dag, M Marsili\ddag, I P\'erez Castillo\P}

\address{\dag\ INFM--SMC and Dipartimento di Fisica, Universit\`a di
Roma ``La Sapienza'', Piazzale Aldo Moro 2, 00185 Roma (Italy)}

\address{\ddag\ The Abdus Salam ICTP, Strada Costiera 11, 34014
Trieste (Italy)}

\address{\P\ Instituut voor Theoretische Fysica, Katholieke
Universiteit Leuven, Celestijnenlaan 200D, 3001 Leuven (Belgium)}

\eads{\mailto{andrea.demartino@roma1.infn.it},
\mailto{marsili@ictp.trieste.it},
\mailto{Isaac.Perez@fys.kuleuven.ac.be}}

\begin{abstract}
The optimal (`equilibrium') macroscopic properties of an economy with
$N$ industries endowed with different technologies, $P$ commodities
and one consumer are derived in the limit $N\to\infty$ with $n=N/P$
fixed using the replica method. When technologies are strictly
inefficient, a phase transition occurs upon increasing $n$. For low
$n$, the system is in an expanding phase characterized by the
existence of many profitable opportunities for new technologies.  For
high $n$, technologies roughly saturate the possible productions and
the economy becomes strongly selective with respect to
innovations. The phase transition and other significant features of
the model are discussed in detail. When the inefficiency assumption is
relaxed, the economy becomes unstable at high $n$.
\end{abstract}

\section{Introduction}

In recent years it has become increasingly clear that the emergence of
non-trivial collective phenomena in systems of heterogeneous
interacting agents (e.g. markets, information networks, or ecosystems)
presents deep analogies with the occurrence of phase transitions in
disordered statistical systems, when heterogeneities in the former can
be mapped onto disorder variables in the latter. The ideas and
techniques of mean-field spin-glass theory \cite{MPV} play a major
role in the current studies of multi-agent systems. As happened in
other interdisciplinary applications of spin-glass theory from neural
networks \cite{hkp} to computer science \cite{mmz}, the effectiveness
of these methods in economics-inspired problems can go well beyond the
purely technical level. Some issues like the emergence of collective
phenomena in evolving populations of competing agents have already
started to be elucidated via spin-glass type of ideas (see e.g. the
literature on the Minority Game \cite{page}). Another intriguing class
of problems arising in economics where such a potential can be fully
expressed concerns the characterization of the equilibria of
heterogeneous economies.

In a nutshell, economic equilibria are optimal microscopic states of
an economy in which (a) all economic agents maximize their respective
payoffs (firms maximize profits and consumers maximize utilities) and
(b) the prices of goods are such that the total demand and the total
supply of each commodity match exactly
\cite{MasColell,Debreu}. Microscopic degrees of freedom are typically
the scale of production of firms and the level of consumption of
consumers, while technologies and endowments or utility functions are
heterogeneous across firms and consumers respectively.  The central
problem is extracting robust macroeconomic properties and laws that
can be tested against observed patterns of real
economies. Heterogeneities complicate things considerably from a
theoretical viewpoint. The traditional approach of economics is based
on the so-called representative agent description, in which each class
of economic agents (firms and consumers) is substituted by a single,
``average'' agent whose characteristics are, to a large extent,
postulated. An alternative pathway consists in modeling
heterogeneities through randomness \cite{Foellmer}. While this
approach has drawn heavily from tools of statistical mechanics, mainly
via analogies with random field models \cite{Durlauf}, the possibility
of systematically employing statistical mechanics to clarify the
statistical structure of economic equilibria has only been partially
explored thus far.

In this work, we apply spin-glass techniques to the study of a random
production economy assuming that technologies (i) are random vectors,
(ii) are to some degree inefficient and (iii) can be operated at any
(non-negative) scale. Within this framework, which is standard in
economics \cite{Lancaster}, it is possible to \emph{derive} an
effective-agent problem whose solution embodies that of the original
multi-agent problem. Key macroeconomic quantities such as averages and
distributions of prices, operation scales and consumptions can be
computed exactly. Remarkably, they turn out to be either directly
connected or easily derived from the spin-glass order parameters that
naturally arise in calculations. Consequently, a macroeconomic picture
emerges naturally from microeconomic interactions and the
corresponding phase structure can be fully characterized.

Our emphasis in this paper will be mainly on `physical' properties of
the model (e.g. phase transitions), so we will not analyze the
behavior of such quantities as the gross domestic product of the
economy. For a more articulated economic discussion, we refer the
reader to \cite{long}, where a slightly simpler model than the one
presented here is studied in a fully economic perspective. The outline
of this work is hence as follows. In Sec. 2 we expound the formal
aspects of the multi-agent problem and outline the spin-glass
approach. Two types of economies are considered: those with strictly
inefficient technologies and those whose technologies are inefficient
\emph{on average}. These two situations give rise to strikingly
different macroscopic properties. The former choice yields a more
realistic economic picture and a non-trivial physical scenario, which
is thoroughly analyzed in Sec. 3. In the latter case, the economy
displays simpler but less realistic properties, that are the subject
of Sec. 4. Finally, in Sec. 5, we fomulate our conclusions and discuss
some possible directions for further work.

\section{The problem and its solution}

\subsection{Statement of the problem}

The definitions of our model parallel those employed in the theory of
economic equilibria, see e.g. \cite{Lancaster}. We consider an economy
with $N$ firms or industries and $P$ commodities. Each industry is
endowed with a technology or activity that allows the transformation
of some commodities (``inputs'') into others
(``outputs''). Technologies are denoted as $P$-dimensional vectors
$\boldsymbol{\xi}_i=\{\xi_i^\mu\}\in\mathbb{R}^P$, where negative
(resp. positive) components represent quantities of inputs
(resp. outputs).  Furthermore, each firm can operate its technology at
a scale $s_i\geq 0$, meaning that when run at scale $s_i$ the industry
$i$ produces or consumes a quantity $\xi_i^\mu s_i$ of commodity
$\mu$. The profit of firm $i$ is
$\pi_i=s_i(\boldsymbol{p}\cdot\boldsymbol{\xi}_i)$, where
$\boldsymbol{p}=\{p^\mu\}$ is the price vector, which we assume to be
non-negative. Each firm chooses $s_i$ by solving
\begin{equation}
\max_{s_i\geq 0}~\pi_i
\label{maxf}
\end{equation}
at fixed prices. Produced goods are bought by consumers to satisfy
their needs. Here we consider the case of a single consumer
(`society') endowed with an initial bundle of goods
$\boldsymbol{y}=\{y^\mu\}$, $y^\mu\geq 0$, and with a prescribed
utility function $U(\boldsymbol{x})$ (`welfare'). The consumer selects
the desired consumption $\boldsymbol{x}=\{x^\mu\}$ by solving
\begin{equation}\label{maxc}
\max_{\boldsymbol{x}\in B}~U(\boldsymbol{x}),
~~~~~~~B=\{\boldsymbol{x}\geq 0 : \boldsymbol{p}\cdot\boldsymbol{x}\le
\boldsymbol{p}\cdot\boldsymbol{y}\}
\end{equation}
at fixed prices. $B$ denotes the consumer's budget set. We specialize
here to the case of a separable utility:
\begin{equation}\label{U}
  U(\boldsymbol{x})=\sum_{\mu=1}^P u(x^\mu)
\end{equation}
which simplifies the analysis considerably. We assume in this way that
commodities are {\em a priori} equivalent for the consumer. While our
calculations hold in general for $u$'s such that $u'(x)>0$ and
$u''(x)<0$, for numerical calculations we set $u(x)=\log x$.

Equilibria are defined as microscopic configurations
$(\boldsymbol{s},\boldsymbol{x})$, with $\boldsymbol{s}=\{s_i\}$ and
$\boldsymbol{x}=\{x^\mu\}$, that solve (\ref{maxf}) and (\ref{maxc})
and such that
\begin{equation}\label{mc}
\boldsymbol{x}=\boldsymbol{y}+\sum_{i=1}^N
s_i\boldsymbol{\xi}_i
\end{equation}
The above ``market-clearing'' condition ensures that the total demand
of each good, $x^\mu-y^\mu$, matches the total supply $\sum_i\xi_i^\mu
s_i$.

We concentrate on the limit of large economies, namely $N\to\infty$,
$P\to\infty$, $n=N/P$ fixed. The consumer's initial endowments $y^\mu$
are assumed to be sampled independently from a distribution $\rho(y)$.
The complexity of the production processes is modeled by postulating
that technologies $\boldsymbol{\xi}_i$ are quenched random vectors. In
order to avoid unrealistic features, it is crucial to impose that it
is not possible to produce, by some combination of the available
technologies, a positive amount of some commodity without consuming a
positive amount of some other commodity (``impossibility of the Land
of Cockaigne'' \cite{Lancaster}). We enforce this condition in two
ways.
\begin{description}
\item[Hard-constrained technologies.] The $\xi_i^\mu$'s are taken to
be Gaussian random variables with zero mean and variance $\Delta_i/P$
satisfying
\begin{equation}
\forall i:\qquad\sum_{\mu=1}^P \xi_i^\mu=-\epsilon_i 
\label{constraint} 
\end{equation}
The parameters $\epsilon_i>0$ measure the inefficiency of the
transformation process of technologies, i.e. the difference between
the total quantity of inputs and the total quantity of outputs.
\item[Soft-constrained technologies.] The $\xi_i^\mu$'s are taken to
be Gaussian random variables with mean $-\epsilon_i/P$ and variance
$\Delta_i/P$.
\end{description}
These two choices give rise to very different macroscopic properties.
In both cases, we further assume that
\begin{equation}\label{eta}
\epsilon_i=\eta\sqrt{\Delta_i}
\end{equation}
and that $\Delta_i$ is drawn from a distribution $g(\Delta)$
independently for each $i$. 

The parameters that ultimately regulate the economy's equilibria are
thus $n$ and $\eta$, together with the functions $u(x)$, $\rho(y)$ and
$g(\Delta)$.

As said above, in this setup commodities are statistically equivalent
in all respects. Hence, this model focuses on the ability of the
economy to produce scarce goods using abundant goods as inputs. Our
aim is thus to understand how the global level of prosperity, measured
by the utility of the consumer and by the amount of economic activity,
increases as the repertoire of available technologies expands.

\subsection{Statistical mechanics approach}

In the following, we shall generally denote by brackets
$\avg{\cdots}_z$ the average over a random variable $z$, irrespective
of its distribution.

To begin with, let us observe that the problem of finding the
equilibrium of the economy ultimately takes the form
\begin{equation}\label{maxu}
\max_{\{s_i\geq 0\}}~
U\l(\boldsymbol{y}+\sum_{i=1}^Ns_i\boldsymbol{\xi}_i\r)
\end{equation}
This is self-evident from the consumer's viewpoint. But if
(\ref{maxu}) is solved, the firms' profit maximization is solved
too. Indeed, taking the partial derivative of $U$ with respect to
$s_i$ we find
\begin{equation}
\frac{\partial U}{\partial s_i}=\sum_{\mu=1}^P\frac{\partial
u}{\partial x^\mu}\xi_i^\mu=\lambda \frac{\partial \pi_i}{\partial
s_i}
\end{equation}
where we used the fact that the consumer's utility maximization,
subject to the budget constraint, implies that $\case{\partial
U}{\partial x^\mu}=\lambda p^\mu$ for some Lagrange multiplier
$\lambda>0$. This is the content of the first Welfare Theorem
\cite{MasColell}: the market achieves an allocation of resources that
is simultaneously optimal for all participants.

Eq.~(\ref{maxu}) is a standard problem in statistical mechanics, which
can be solved by introducing a fictitious temperature, building a
partition function with $-U$ in place of the Hamiltonian, and then
letting the temperature to zero. In order to cope with the disorder
(either hard- or soft-constrained), it is necessary to introduce
replicas before averaging. This standard procedure generates
macroscopic order parameters such as
\begin{equation}\label{defQ}
  q_{ab}=\frac{1}{N}\sum_{i=1}^N\Delta_i s_{ia} s_{ib}
\end{equation}
where the indices $a,b=1,\ldots,r$ run over replicas. The convexity
properties of $U$ ensure that the solution of the maximization problem
is unique. This implies that the replicated partition function, which
in the limit $N\to\infty$ can be computed via a simple saddle-point
integration, is dominated by replica-symmetric saddles where
\begin{equation}
q_{a,b}=Q\delta_{a,b}+q(1-\delta_{a,b}).
\end{equation}
(in spin-glass jargon, $Q$ represents the self-overlap while $q$ is
the analog of the Edwards-Anderson order parameter). Here, we will
leave routine calculations aside (details of a very similar
calculation can be found in \cite{long}). Let it suffice to say that
after taking the thermodynamic limit $N\to\infty$, the replica limit
$r\to 0$, and the limit of zero temperature, one arrives at an
expression of the form
\begin{equation}\label{maxures}
\lim_{N\to\infty}\frac{1}{P}\avg{\max_{\{s_i\geq 0\}}~
U\left(\boldsymbol{y}+\sum_{i=1}^Ns_i\boldsymbol{\xi}_i
\right)}_{\boldsymbol{\xi}}= {\rm
extr}_{\boldsymbol{\omega}}~f(\boldsymbol{\omega})
\end{equation}
where $\boldsymbol{\omega}$ is a vector of order parameters and ${\rm
extr}$ means that the solution is provided by the saddle point of the
function $f$, i.e. by the vector $\boldsymbol{\omega}^*$ that solves
the equation $\case{\partial f}{\partial\boldsymbol{\omega}}=0$. The
function $f$ depends in this case on the type of disorder
$\boldsymbol{\xi}$ one is considering.

We shall first focus on the case of hard-constrained technologies, for
which a well-behaved solution exists for all $n$, as long as
$\eta>0$. Such a solution exhibits two regimes, separated by a phase
transition in the limit $\eta\to 0$. For small values of $n$, when few
technologies are available, the economy possesses many unexploited
productive opportunities and the introduction of new technologies
increases the productivity of existing ones. For larger values of $n$,
the economy enters a mature phase, where the productive sector is
largely saturated by existing technologies.

In the soft-constrained case, instead, the economy becomes unstable as
the average scale of production diverges for $n$ larger than a
critical $\eta$-dependent threshold. We will show explicitly that such
an instability is related to the fact that, above this threshold, it
is possible (with probability one) to combine existing technologies in
order to produce outputs without using inputs.

\section{Hard-constrained technologies}

\subsection{The effective-agent problems}

If all technologies are strictly inefficient (i.e. if, for each $i$,
$\sum_\mu\xi_i^\mu=-\eta\sqrt{\Delta_i}$), one finds that equilibria
are described by the saddle point of
\begin{eqnarray}\label{effe}
\fl f(Q,\gamma,\chi,\widehat{\chi},\kappa,p)
=\frac{1}{2}nQ\widehat{\chi} -\frac{1}{2}\gamma\chi+\kappa
p\nonumber\\ +\avg{\max_{x\geq 0}
\l[u(x)-\frac{1}{2\chi}\l(x-y+t\sqrt{nQ}+\kappa\r)^2\r]}_{t,y}+\nonumber\\
+n\avg{\max_{s\geq 0}\l[-\frac{1}{2}\Delta\widehat{\chi}s^2+
st\sqrt{\Delta(\gamma-p^2)}-s\eta p\sqrt{\Delta}\r]}_{t,\Delta}
\end{eqnarray}
where $t$ is a unit Gaussian random variable, and averages over $y$
and $\Delta$ are performed with distributions $\rho(y)$ and
$g(\Delta)$, respectively. The representative-agent problem embodying
the equilibrium structure of the multi-agent economy is rather
clear. The first maximization problem on the r.h.s. can be interpreted
as an effective-utility maximization by the consumer with respect to
the consumption of a single ``representative'' commodity. The second
one corresponds instead to effective-profit maximization by a
``representative'' company with respect to its operation scale. The
two problems are interconnected in a non-trivial way by the other
terms. The solutions $x^*\equiv x^*(t,y)$ and $s^*\equiv
s^*(t,\Delta)$ of the effective problems as functions of the random
variables $(t,y)$ and $(t,\Delta)$, respectively, are given by
\begin{equation}\label{xstar}
x^*\equiv x^*(t,y)~~~~{\rm such~that}~~~~ \chi
u'(x^*)=x^*-y+t\sqrt{nQ}+\kappa
\end{equation}
and
\begin{equation}\label{sstar}
s^*(t,\Delta)=\cases{\frac{t\sigma-\eta
p}{\widehat{\chi}\sqrt{\Delta}} & for $t\geq \eta p/\sigma$\\ 0&
otherwise}
\end{equation}
Notice that the equilibrium consumption $x^*$ is non-negative as long
as $u'(x)$ diverges when $x\to 0$. Notice also that the dependence of
the equilibrium scale $s^*$ on $\Delta$ is described by
\begin{equation}\label{scaling}
s^*(t,\Delta)=\frac{1}{\sqrt{\Delta}}~s^*(t,1)
\end{equation}
Moreover, by introducing the re-scaled variable
$\widetilde{s}=s\sqrt{\Delta}$ the dependence of (\ref{effe}) on
$\Delta$ disappears.  This simple scaling will greatly simplify our
discussion of the effects induced by heterogeneities in $\Delta$ (next
section).

Using the shorthand $\sigma=\sqrt{\gamma-p^2}$ and a little algebra,
the saddle-point conditions can be seen to take the following form:
\begin{eqnarray}
p=\avg{u'(x^*)}_{t,y}\label{p}\\
\sigma=\sqrt{\l[\avg{u'(x^*)^2}_{t,y}-\avg{u'(x^*)}^2_{t,y}\r]}\label{sigma}\\
\widehat{\chi}=\frac{1}{\sqrt{nQ}}\avg{t
u'(x^*)}_{t,y}\label{chihat}\\
Q=\avg{(s^*)^2\Delta}_{t,\Delta}\label{Q}\\
\chi=\frac{n}{\sigma}\avg{t s^*\sqrt{\Delta}}_{t,\Delta} \label{chi}\\
\kappa=p\chi+n\eta\avg{s^*\sqrt{\Delta}}_{t,\Delta}\label{kappa}
\end{eqnarray}
In the next sections we shall discuss the generic properties of the
above saddle-point problem and the corresponding phase structure.

\subsection{Economic interpretation}

The meaning of the macroscopic order parameters and of the
saddle-point conditions can be easily found with at most some
basic algebra.
\begin{itemize}
\item Equation (\ref{p}) implies that the order parameter $p$ is the
average (relative) price, while $\sigma$ (see (\ref{sigma})) measures
(relative) price fluctuations. This is so because utility maximization
under budget constraint gives $\case{\partial u}{\partial
x^\mu}=\lambda p^\mu$ with $\lambda>0$ a Lagrange multiplier that can
be set to $1$ without any loss of generality. Evaluating this at the
saddle point gives precisely (\ref{p}).
\item Averaging the expression in (\ref{xstar}) over $t$ and $y$ and
expressing $\kappa$ via (\ref{kappa}) one finds
\begin{equation}\label{mcra}
\avg{x^*}_{t,y}=\avg{y}_y-n\eta\avg{s^*\sqrt{\Delta}}_{t,\Delta}
\end{equation}
This equation can also be obtained by averaging the market clearing
condition (\ref{mc}) over technologies taking (\ref{constraint}) into
account. Hence (\ref{mcra}) can be seen as the macroeconomic analog of
(\ref{mc}).
\item It is possible to show via elementary manipulations \cite{long}
that
\begin{equation}
\avg{u'(x^*)(x^*-y)}_{t,y}=0
\end{equation}
Remembering that $u'(x^*)$ is the relative price at equilibrium, this
equation implies that the consumer saturates his/her budget when
choosing his/her consumption. This law, known in economics as Walras'
law \cite{MasColell}, is a generic consequence of the microscopic
setup of economic equilibria. The fact that it can be retrieved from
our macroscopic saddle-point problem constitutes a significant
consistency check.
\end{itemize}

\subsection{Equilibrium distributions of operation scales, consumptions 
and (relative) prices}

The distribution of operation scales $\mathcal{P}(s)$ provides
important information, for it can be seen as a proxy for the
distribution of firm sizes (either in terms of revenues or in terms of
the number of employees). Its calculation in our model can be carried
out with the help of (\ref{scaling}). In fact, one has
\begin{equation}
\mathcal{P}(s)=\int_0^\infty g(\Delta) P(s|\Delta) d\Delta
\end{equation}
where $P(s|\Delta)$ is the distribution of operation scales at fixed
$\Delta$. The latter can be derived from (\ref{sstar}) and from the
(Gaussian) distribution of $t$:
\begin{equation}
\fl P(s|\Delta)=\avg{\delta(s-s^*(t,\Delta))}_t=(1-\phi)\delta(s)+
\frac{\widehat{\chi}}{\sqrt{2\pi\sigma^2}}~e^{-\frac{(\widehat{\chi}s
\sqrt{\Delta}+\eta p)^2}{2\sigma^2}}\theta(s)
\end{equation}
with
\begin{equation}\label{phi}
\phi=\avg{\theta\l(t-\frac{\eta
p}{\sigma}\r)}_t=\frac{1}{2}\l[1-\erf\l(\frac{\eta
p}{\sigma\sqrt{2}}\r)\r]
\end{equation}
the fraction of active firms (i.e. firms such that $s_i>0$), which is
independent of $\Delta$. Now (\ref{scaling}) implies that
$P(s|\Delta)=\sqrt{\Delta} P(s\sqrt{\Delta}|1)$ so that
\begin{equation}
\mathcal{P}(s)=\frac{2}{s^3}\int_0^\infty k^2 g\l(\frac{k^2}{s^2}\r)
P(k|1) dk
\end{equation}
where we simply set $k=s\sqrt{\Delta}$. The behavior of $Q(s)$ for
large $s$ is related to the behavior of $g(\Delta)$ for small
$\Delta$. In particular, if $g(\Delta)\propto \Delta^\gamma$ for
$\Delta\ll 1$, then $\mathcal{P}(s)\propto s^{-3-2\gamma}$ for $s\gg
1$, which agrees with the power law behavior found empirically for the
distribution of firm sizes \cite{japan,usa}.

The distribution of equilibrium consumptions at fixed endowments can
be computed from (\ref{xstar}) via the distribution of $t$. One gets
\begin{equation}
P(x|y)=\avg{\delta(x-x^*(t,y))}_t=\frac{1-\chi u''(x)}{\sqrt{2\pi n
Q}}~e^{-\frac{(x-y-\chi u'(x)+\kappa)^2}{2nQ}}
\end{equation}
As before, the equilibrium distribution of consumptions is given by
\begin{equation}
\mathcal{P}(x)=\int_0^\infty \rho(y) P(x|y) dy
\end{equation}

Likewise, it is also possible to compute, via the simple change of
variable $p=u'(x)$, the equilibrium distribution of prices.

\subsection{Limit $\rho(y)\to\delta(y-\ovl{y})$}

If the spread of initial endowments becomes vanishingly small, the
amount of productive activity is expected, quite intuitively, to
vanish too. In order to see this, one can proceed as follows. Let the
initial endowment be expressed as
\begin{equation}
y=\ovl{y}+\zeta
\end{equation}
with $\zeta$ a (small) r.v. with zero average and $\ovl{y}$ a positive
number, and let
\begin{equation}
x^*(t,y)\equiv \ovl{y}+\psi ~~~~~~~ \psi\equiv\psi(t,\zeta)
\end{equation}
Anticipating that $\psi\sim\zeta$ will be a small quantity when
$\avg{\zeta^2}_\zeta\ll 1$, we can write to leading order
\begin{equation}\label{zazza}
u'(x^*)\simeq u'(\ovl{y})+\psi u''(\ovl{y}).
\end{equation}
Likewise,
\begin{eqnarray}
p\simeq u'(\ovl{y})+u''(\ovl{y})\avg{\psi}_{t,\zeta}\label{da}\\
\widehat{\chi}\simeq\frac{u''(\ovl{y})}{\sqrt{nQ}}\avg{t\psi}_{t,\zeta}
\label{db}\\ \sigma^2\simeq [u''(\ovl{y})]^2\l(\avg{\psi^2}_{t,\zeta}-
\avg{\psi}^2_{t,\zeta}\r)\label{dc}
\end{eqnarray}
follow immediately from (\ref{p}), (\ref{sigma}) and (\ref{chihat}).
An expression for $\psi$ can be obtained from (\ref{xstar}) via
(\ref{zazza}):
\begin{equation}\label{equazia}
\chi[u'(\ovl{y})+\psi u''(\ovl{y})]\simeq\kappa+\psi-\zeta+t\sqrt{nQ}
\end{equation}
Separating terms of order $\psi^0$ from those proportional to $\psi$,
one has
\begin{eqnarray}
\kappa=\chi u'(\ovl{y})\\ 
\psi(t,\zeta)=\frac{\zeta-t\sqrt{nQ}}{1-\chi
u''(\ovl{y})}\label{psi}
\end{eqnarray}
Substituting (\ref{psi}) into (\ref{da}), (\ref{db}) and (\ref{dc})
and carrying out the averages one gets
\begin{eqnarray}
p\simeq u'(\ovl{y})\\ \widehat{\chi}\simeq -\frac{u''(\ovl{y})}{1-\chi
u''(\ovl{y})}\label{nchihat}\\
\sigma^2\simeq \widehat{\chi}^2\l(\avg{\zeta^2}_\zeta+nQ\r)
\label{nsigma}
\end{eqnarray}
In order to calculate $Q$, we can use (\ref{nsigma}). Insertion into
(\ref{sstar}) yields $s^*=(t-\tau)s_0\theta(t-\tau)$, where
\begin{eqnarray}\label{namo}
s_0=\frac{\sigma}{\widehat{\chi}\sqrt{\Delta}}\simeq
\sqrt{\frac{\avg{\zeta^2}_\zeta+nQ}{\Delta}}\\ \tau=\frac{\eta
p}{\sigma}\simeq\frac{\eta p}{\widehat{\chi}}\frac{1}{\sqrt{
\avg{\zeta^2}_\zeta +nQ}}\label{rinamo}
\end{eqnarray}
In turn, from (\ref{Q}) we obtain
\begin{equation}\label{nQ}
Q\simeq \l(\avg{\zeta^2}_\zeta+nQ\r)F(\tau)~~~\Rightarrow
~~~Q\simeq\frac{\avg{\zeta^2}_\zeta F(\tau)}{1-n F(\tau)}
\end{equation}
where
\begin{equation}\label{nQF}
F(\tau)=\avg{(t-\tau)^2\theta(t-\tau)}_t
\end{equation}
Now (\ref{nQ}) and (\ref{nQF}) together imply that $\tau\sim
-\avg{\zeta^2}_\zeta^{-1/2}$ for small $\avg{\zeta^2}_\zeta$ so that,
when $\avg{\zeta^2}_\zeta\to 0$, one can resort to an asymptotic
expansion of $F(\tau)$ for $\tau\gg 1$ to calculate $Q$. This gives
\begin{equation}
Q\simeq \avg{\zeta^2}_\zeta^{3/2} \exp\l[-\frac{\eta^2 [u'(\ovl{y})]^2}{
2[u''(\ovl{y})]^2\avg{\zeta^2}_\zeta}\r]
\end{equation}
So $Q\to 0$ for $\avg{\zeta^2}_\zeta\to 0$ with an essential
singularity. In turn, by virtue of (\ref{nsigma}) and (\ref{phi}),
respectively, $\sigma\to 0$ and $\phi\to 0$ for
$\avg{\zeta^2}_\zeta\to 0$. Hence when the spread of initial
endowments decreases, economic activity vanishes very rapidly and no
market activity takes place when initial endowments are all equal.

\subsection{Limit $\eta\to 0$: the phase transition}

Before discussing the actual solutions of the saddle-point equations
as a function of $n$ and the corresponding economic picture, let us
briefly investigate the limit $\eta\to 0$ of marginally efficient
technologies, in which the critical properties of the model can be
elucidated in detail. When $\eta=0$, one sees from (\ref{phi}) that
the fraction of active companies becomes $\phi=1/2$, while from
(\ref{sstar}) one gets
\begin{equation}
s^*(t,\Delta)=\cases{\frac{t\sigma}{\widehat{\chi}\sqrt{\Delta}} & for
$t\geq 0$\\ 0& for $t\leq 0$}
\end{equation}
The saddle-point conditions (\ref{Q}), (\ref{chi}) and (\ref{kappa})
in turn take the simple form
\begin{equation}\label{erma}
Q=\frac{\sigma^2}{2\widehat{\chi}^2}~~~~~~~
\chi=\frac{n}{2\widehat{\chi}}~~~~~~~
\kappa=p\chi
\end{equation}
One sees immediately that this provides a solution that must break
down for $n>2$, as there cannot be more than $P$ active firms in the
economy (i.e. $\phi$ must satisfy the condition $n\phi\le 1$). Let us
then focus on the limit $n\to 2^-$ by studying the case $0<2-n\ll
1$. When there is an almost complete set of technologies (i.e. when
$n\phi\simeq 1$), goods can be transformed in almost all ways. Hence
we expect that in equilibrium $x^*=\ovl{y}+\psi$ with
$\psi\equiv\psi(t)$ small. Then we can again use (\ref{zazza}) and
(\ref{xstar}) to obtain an expression for $\psi$ (as done in
(\ref{equazia})). After simple algebra we find
\begin{equation}
\psi=\frac{2\widehat{\chi}(\ovl{y}-y)+t\sigma\sqrt{2n}}{n
u''(\ovl{y})-2\widehat{\chi}}
\end{equation}
Substituting this into (\ref{nchihat}) and (\ref{nsigma}) one obtains
\begin{eqnarray}
\widehat{\chi}=-\frac{1}{2}u''(\ovl{y})(2-n)\\
\sigma^2=\frac{1}{2}[u''(\ovl{y})]^2(2-n)
\end{eqnarray}
and thus, while the fluctuation of relative prices $\sigma\to 0$,
because of (\ref{erma}), $Q\sim \case{1}{2-n}$ and $\chi\sim
\case{1}{2-n}$. This means that the typical scale of production
diverges as $\case{1}{\sqrt{2-n}}$ when $n\to 2^-$.

A slightly more refined analysis is required to investigate the limit
$\eta\to 0$ for $n>2$. To simplify equations, for the remainder of
this section we fix $\Delta=1$ and resort to the notation introduced
in the previous section, so that $s^*=(t-\tau)s_0\theta(t-\tau)$ with
$s_0$ and $\tau$ as defined in the equalities in (\ref{namo}) and
(\ref{rinamo}). Again anticipating that, because the set of
technologies is almost complete and efficient, the equilibrium
consumption levels $x^*=\ovl{y}+\psi$ will fluctuate very little
around the average $\ovl{y}$ (i.e. $\psi\ll 1$), we can look for
solutions of the saddle-point problem such that $\sigma$ vanishes
linearly with $\eta$ (i.e. such that $\sigma\sim c\eta$) and with
$s_0=\sigma/\widehat{\chi}$ finite (i.e. such that $\widehat{\chi}$
vanishes linearly with $\eta$ as well). In doing so, we will retain
only the leading terms in $\eta$ for the rest of this section.  Now
$\tau=p/c$, so that from (\ref{Q}) we have
\begin{equation}\label{baba}
Q=s_0^2 F(p/c)
\end{equation}
with $F(\cdot)$ as in (\ref{nQF}), whereas from (\ref{chi}) one gets
\begin{equation}\label{abab}
\chi=\frac{n s_0}{c\eta}H(p/c)
~~~~~~~ H(x)=\frac{1}{2}\l[1-\erf\l(\frac{x}{\sqrt{2}}\r)\r]
\end{equation}
Expanding (\ref{xstar}) as in (\ref{zazza}), and using (\ref{baba})
and (\ref{abab}), one obtains
\begin{equation}
\psi\simeq\frac{c\eta}{n|u''(\ovl{y})|H(p/c)}
\l[\frac{1}{s_0}(y-\ovl{y}) -t\sqrt{n F(p/c)}\r]
\end{equation}
Inserting the above and (\ref{baba}) into (\ref{chihat}) one finds
\begin{equation}
n^2 H(p/c)^2\simeq 1
\end{equation}
This, via (\ref{phi}), confirms that $H(p/c)\equiv\phi\simeq 1/n$
(i.e. that technologies saturate the possible productions). Moreover,
as $H(x)< 1/2$ for $x>0$, it allows us to conclude that this solution
holds for $n>2$. Finally, it tells us that $c\to\infty$ for $n\to 2^+$
as $c\sim\case{1}{n-2}$.

It remains to evaluate $s_0$. To this aim, we use (\ref{sigma}). After
minor algebraic manipulations, it yields
\begin{equation}
s_0\sim\frac{1}{\sqrt{1-nF(p/c)}}
\end{equation}
For $n\to 2^+$ one has $c\to\infty$ and hence $F(p/c)\to 1/2$. So the
typical scale of production (which is proportional to $s_0$) diverges
as $\case{1}{\sqrt{n-2}}$ for $n\to 2^+$.

In summary, we found that for $\eta\to 0$ a critical value $n_c=2$ of
$n$ exists such that
\begin{equation}\label{phiphi}
\phi= \cases{1/2 & for $n<n_c$\\ 1/n & for $n>n_c$}
\end{equation}
while the typical scale of operations behaves as
\begin{equation}\label{esseesse}
\avg{s^*}\sim\frac{1}{\sqrt{|n-n_c|}} ~~~~~~~ |n-n_c|\ll 1
\end{equation}
This shows that the behavior of the economy close $n=2$ for $\eta\ll
1$ is totally analogous to that of a second-order phase transition in
statistical physics.

\subsection{Transition to the Land of Cockaigne}

\begin{figure}
\begin{center}
\includegraphics[width=7cm]{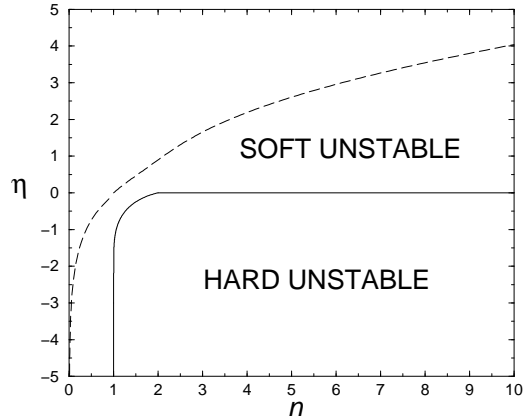}
\caption{\label{pd}Phase diagram. Critical lines $n_c(\eta)$ for the
economy with hard-constrained technologies (solid line) and
soft-constrained technologies (dashed line). The economies are
unstable for $n>n_c(\eta)$.}
\end{center}
\end{figure}

The analysis performed so far can be easily extended to negative
values of $\eta$. In this case, it is to be expected that the economy
becomes unstable at certain critical points $n_c(\eta)$. The reason is
that, for each fixed $\eta<0$, combinations of technologies allowing
to produce all goods without absorbing inputs (i.e. such that $\sum_i
s_i\xi_i^\mu>0$ for all $\mu$) become possible with probability one
when $n$ is sufficiently large, larger than a critical
$\eta$-dependent value $n_c(\eta)$. The equilibrium is stable for
$n<n_c(\eta)$, whereas equilibrium operation scales diverge as $n\to
n_c(\eta)$. Once this has happened, initial endowments become
irrelevant, so that $n_c(\eta)$ is independent of $\rho(y)$. Rather,
it only depends on the asymptotic behavior of the utility function
$u(x)$. Fig.~\ref{pd} shows the critical line $n_c(\eta)$ (solid line)
separating the stable from the unstable phase for $u(x)=\log x$ and
$g(\Delta)=\delta(\Delta -1)$. This line was obtained by imposing the
condition $n\phi=1$ on the solution of the saddle point equations with
$s_0\to\infty$. As $\eta$ decreases, $n_c(\eta)$ rapidly approaches
$1$, with an exponential behavior.

Notice that while this transition is continuous and characterized by
divergences (i.e. $s_i\to\infty$) for $\eta\le 0$ as $n\to n_c(\eta)$,
it is discontinuous across the $\eta=0$ line for $n>2$. Indeed all
scales of consumption $\avg{x^*}$ and production $\avg{s^*}$ remain
finite as $\eta\to 0^+$ for $n>2$ and diverge abruptly as soon as
$\eta<0$. This makes the phase diagram of this model very similar to
that of Minority Games with market-impact correction
\cite{DeMartinoMarsili}. Broadly speaking, both models deal with a
system where a number $N$ of agents compete for the exploitation of a
fixed number $P$ of resources and in both cases $\eta$ tunes the
efficiency of agent's behavior.

\subsection{Solution in a typical example}

Let us finally discuss the solutions of the saddle-point problem as a
function of $n$. We have chosen, for simplicity, the following
parameters:
\begin{equation}\label{parama}
g(\Delta)=\delta(\Delta-1)~~~~~~~ u(x)=\log x ~~~~~~~ \rho(y)=e^{-y}
\end{equation}
Correspondingly, we have obtained numerically the solution of Eq.s
(\ref{p}--\ref{kappa}) as a function of $n$ for different values of
$\eta$. The resulting economic observables, representing equilibrium
values, have been compared to those of equilibria computed numerically
via a simple gradient descent algorithm for a toy economy with $P=16$
and $P=32$ commodities.

\begin{figure}
\begin{center}
\includegraphics[width=9cm]{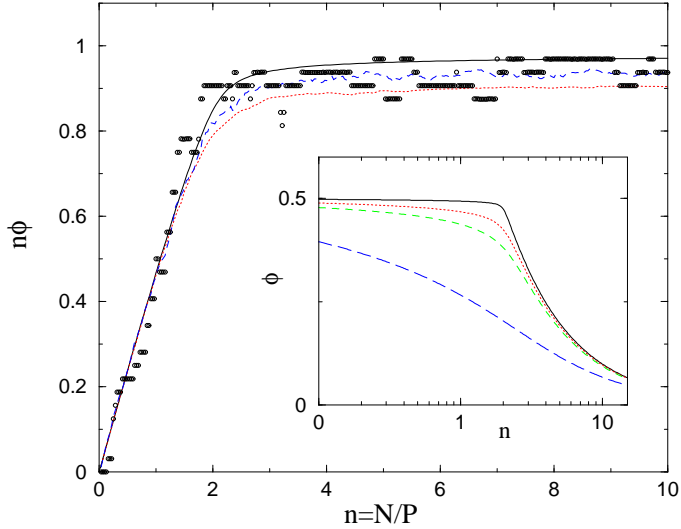}
\caption{\label{fig-phi}Economy with hard-constrained
technologies. Behaviour of $n\phi$ ($\phi=$ fraction of active
companies) at equilibrium as a function of $n$ for $\eta=0.05$ and
$g(\Delta)=\delta(\Delta-1)$: analytical prediction (continuous line),
computer experiments with $P=16$ (dotted line) and for $P=32$ (dashed
line) averaged over 100 disorder samples. Dots represent results of a
single realization of the technologies. Inset: $\phi$ vs $n$ for
$\eta=0.01, 0.05, 0.1, 0.5$ (top to bottom)}
\end{center}
\end{figure}

The fraction of active firms $\phi$ (or $n\phi$) is shown in
Fig.~\ref{fig-phi}.
The two regimes described by (\ref{phiphi}) appear distinctly for low
$\eta$. For $n<n_c=2$, $\phi$ is roughly equal to $1/2$. Here, about a
half of the existing technologies are profitable. For $n>n_c$,
instead, $\phi\simeq 1/n$, i.e. the number of active firms saturates
the number of goods and the economy becomes extremely selective. This
picture persists for larger values of $\eta$, although the transition
appears to be more and more blurred as $\eta$ increases.

\begin{figure}
\begin{center}
\subfigure{\scalebox{.33}{\includegraphics{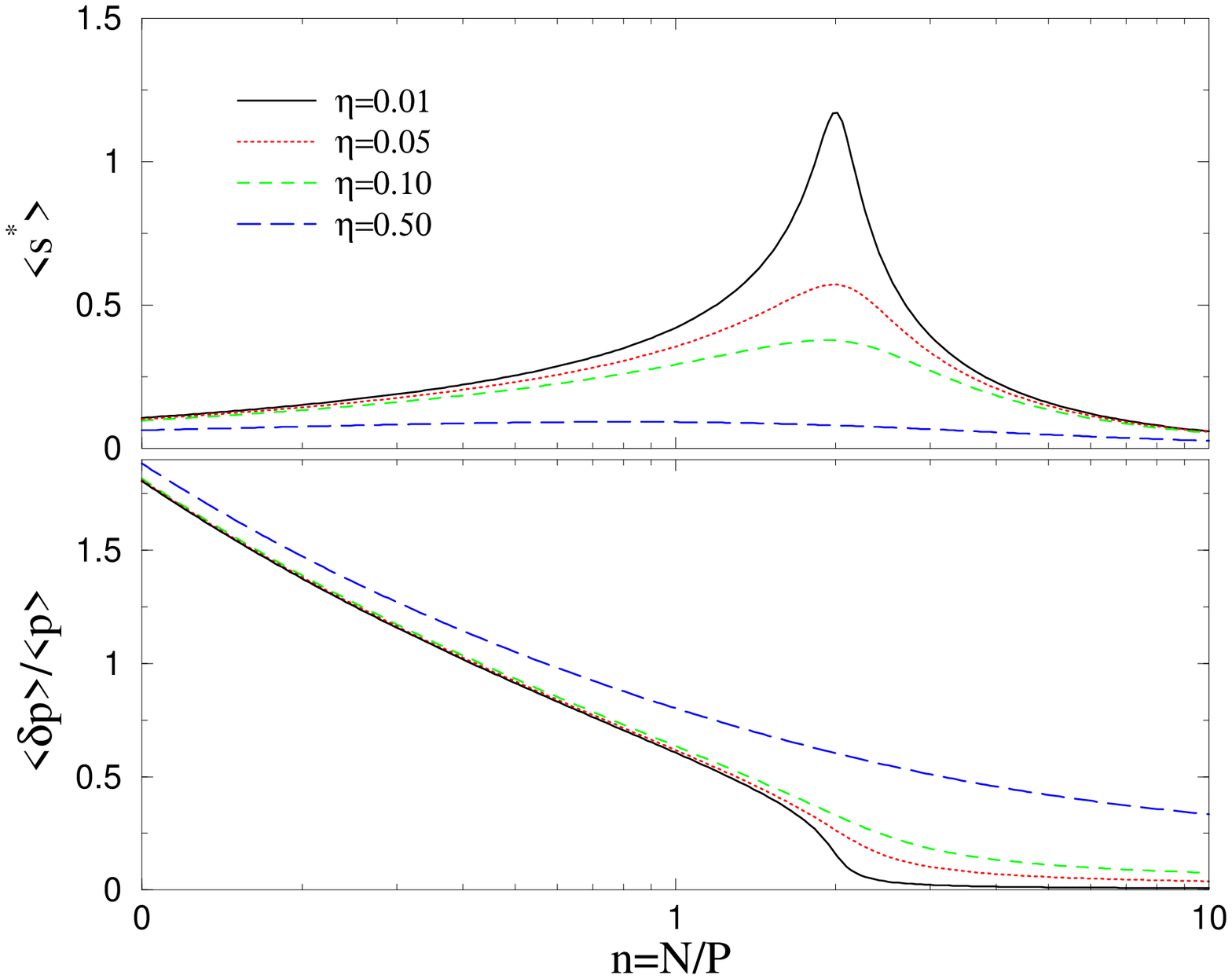}}}
\subfigure{\scalebox{.34}{\includegraphics{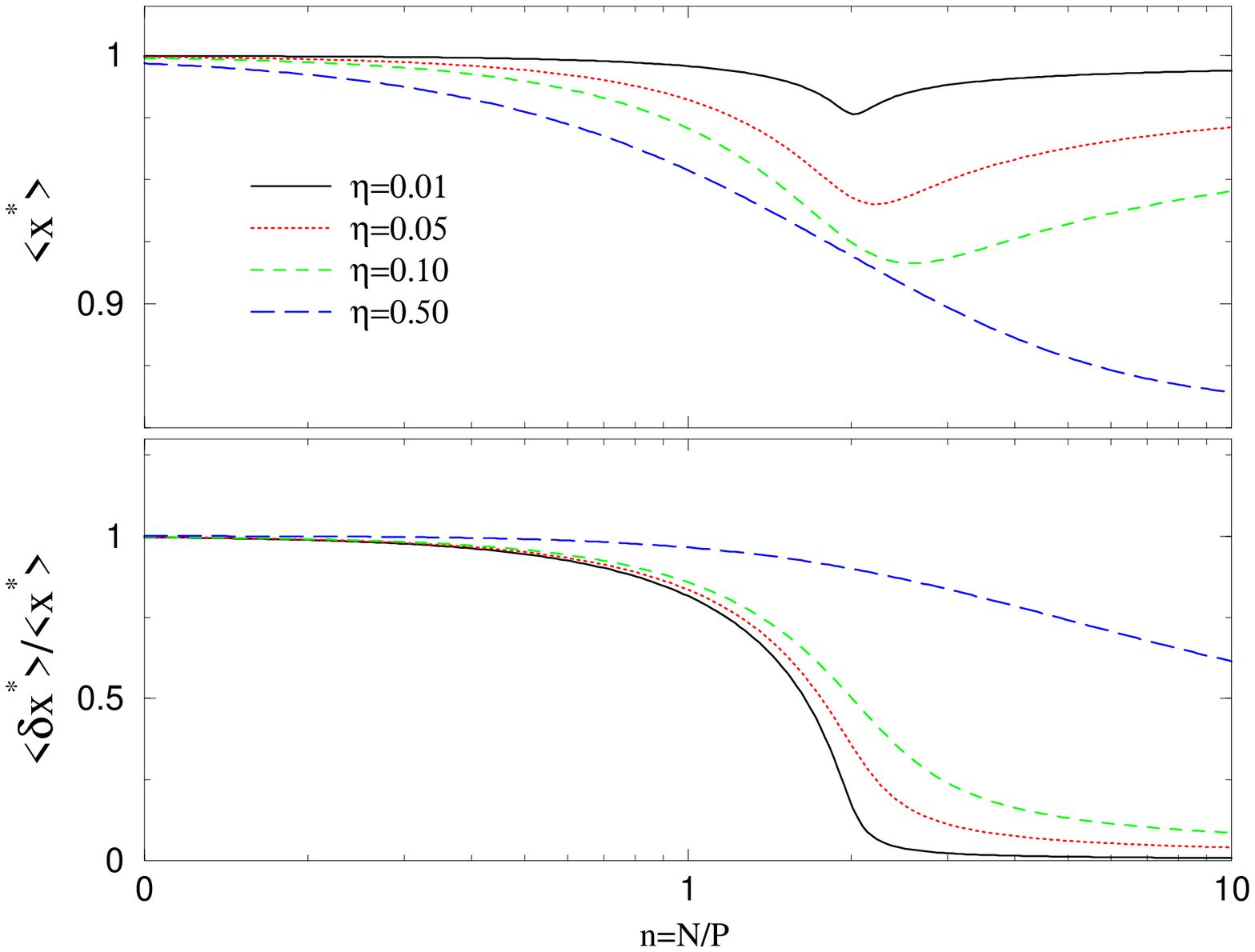}}}
\caption{\label{fig-sx}Economy with hard-constrained
technologies. Behaviour of some relevant economic observables at
equilibrium with $g(\Delta)=\delta(\Delta-1)$. Top left panel: average
scale of operation. Bottom left panel: relative price
fluctuations. Top right panel: average consumption level. Bottom right
panel: relative fluctuations of the consumption level. Values of
$\eta$ are as indicated.}
\end{center}
\end{figure}

In Fig.~\ref{fig-sx} we display the equilibrium behaviour of some
economically relevant quantities. Let us discuss the behavior for
$n<n_c=2$ first. One sees that the average scale of production
increases when $n$ grows. This can be interpreted by saying that all
firms benefit from the introduction of new technologies (i.e. from an
increase of $n$) if the market is not too selective. In parallel,
relative price fluctuations decrease as $n$ increases, as does the
average level of consumption, signalling that firms are managing the
transformation of abundant goods into scarce ones. The distribution of
consumptions preserves the character of the original distribution of
endowments for low $n$ \cite{long}, while relative fluctuations stay
roughly unchanged.

When $n$ is close to $n_c$, operation scales become larger and larger
as $\eta$ decreases (i.e. as technologies become more and more
efficient) and ultimately develop a singularity at $n_c$ in the
marginally efficient limit $\eta\to 0$ (see (\ref{esseesse})). The
fluctuations of relative consumptions start to drop (the sharper the
lower is $\eta$), as the distribution of consumptions becomes more and
more peaked around the mean value. Identifying abundant (or scarce)
goods becomes increasingly hard.

In the saturated regime $n>n_c$, technological innovations ($N\to
N+1$) lead to a decrease in the average operation scale, i.e. new
profitable technologies punish existing ones. Firms cannot take
advantage of the spread between scarce and abundant goods any longer,
and technologies are subject to a much stronger selection. On the
other hand, the average consumption starts growing with $n$, as is
expected in a competitive economy that selects highly efficient
technologies.

It is interesting to notice (see \cite{long} for a more detailed
economic discussion) that while technological innovations
(i.e. increases of $N$ at fixed $P$) improve the global economic
picture for $n<n_c$, when $n>n_c$ a global improvement requires
broadening the range of disposable commodities (i.e. an increase of
$P$ at fixed $N$, or a decrease of $n$). Within the obvious
limitations of this model, this seems to suggest that an economy
driven by technological innovation and products creation may evolve
`spontaneously' toward the critical point $n_c$ under its agents'
maximizing pressures.

\section{Soft-constrained technologies}

\subsection{The effective-agent problem and its typical solutions}

\begin{figure}
\begin{center}
\subfigure{\scalebox{.4}{\includegraphics{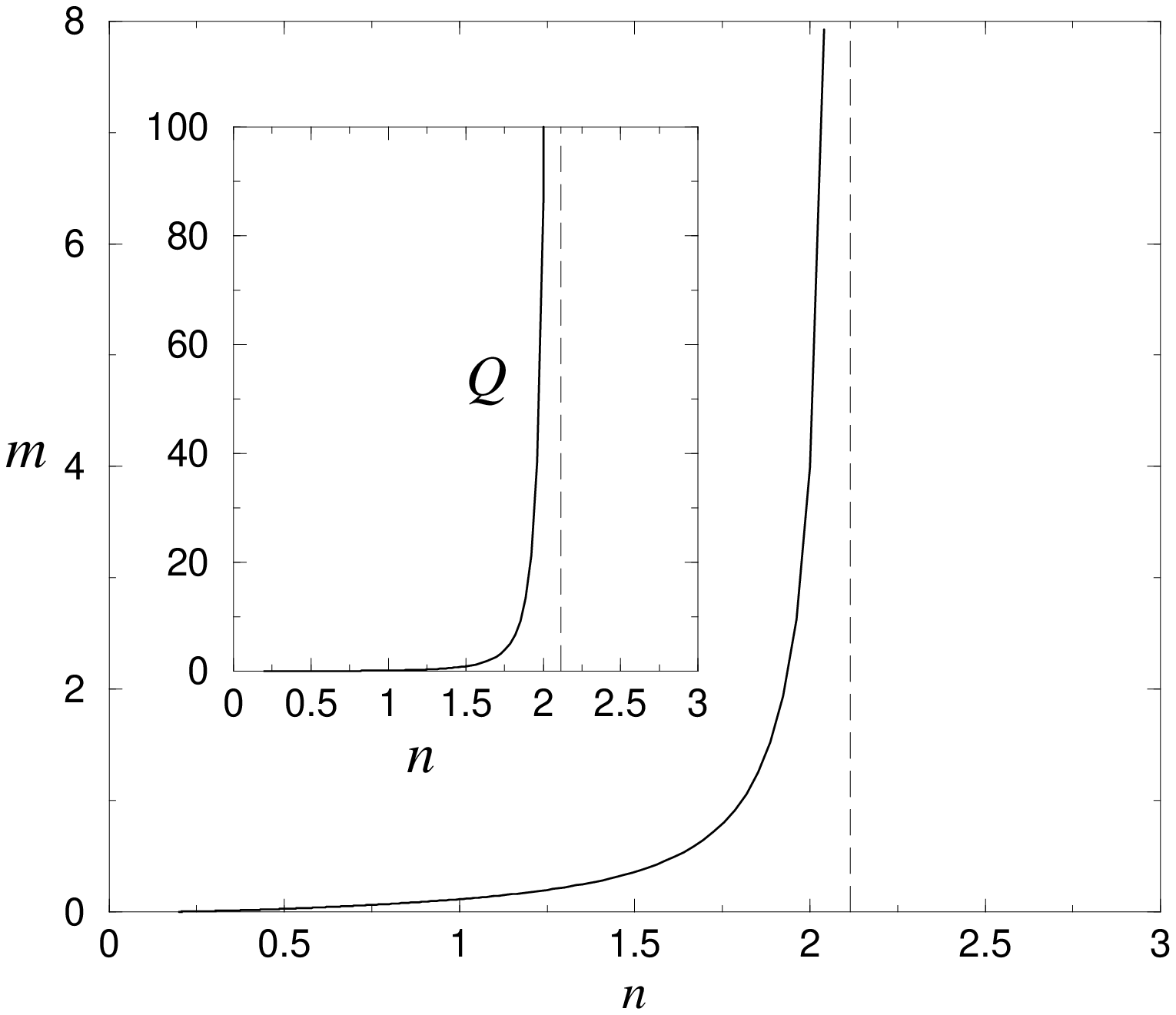}}}
\subfigure{\scalebox{.4}{\includegraphics{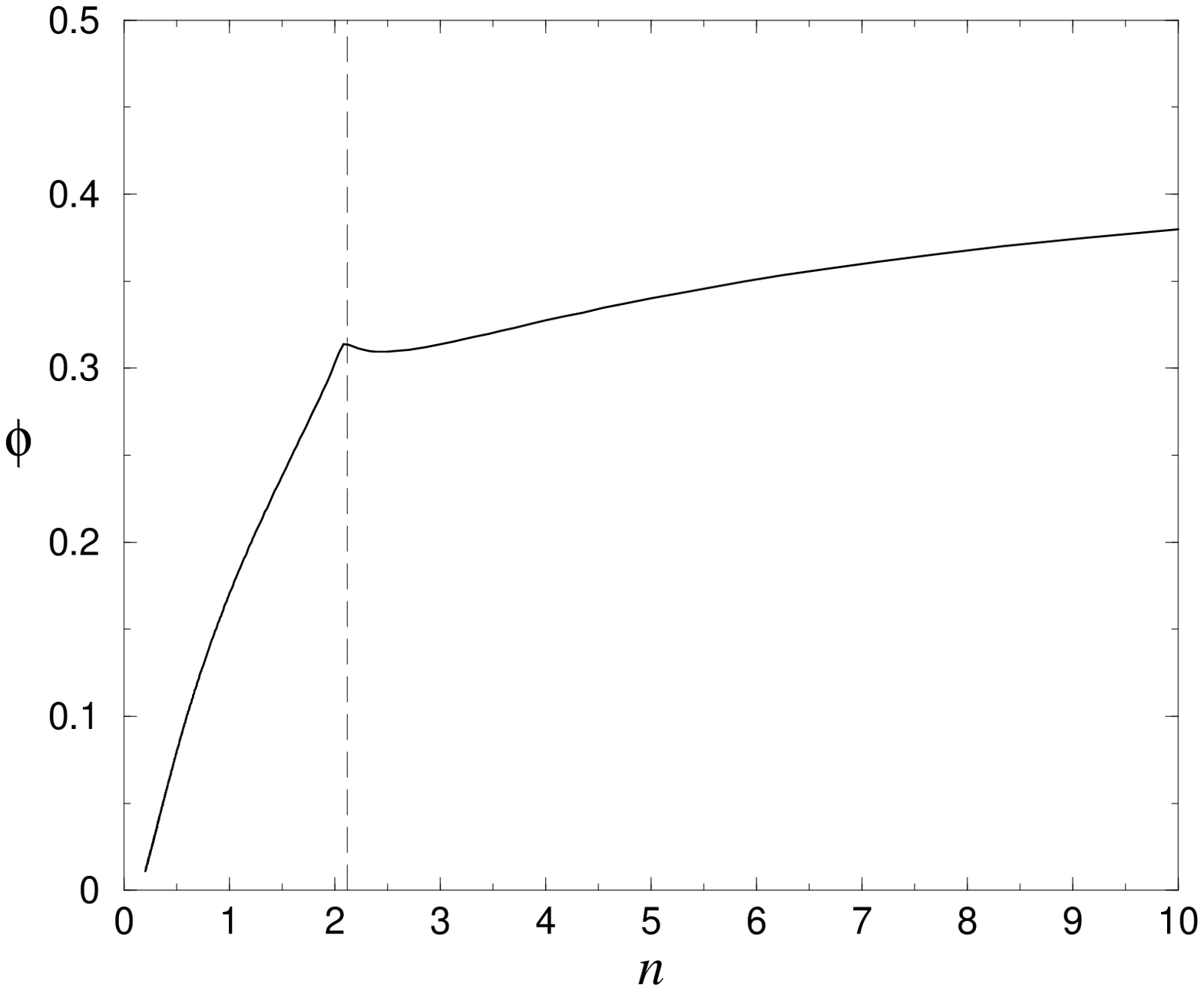}}}
\caption{\label{mQphi}Economy with soft-constrained
technologies. Solution of the saddle-point equations as a function of
$n$ for $\eta=1$ and $g(\Delta)=\delta(\Delta-1)$. Left panel: average
scale of operation $m$ and self-overlap $Q$. Right panel: fraction of
active companies. The dashed vertical lines mark the position of the
critical point $n_c$ calculated analytically.}
\end{center}
\end{figure}

Let us now turn our attention to the case of soft-constrained
technologies. The analog of (\ref{effe}) in this case is
\begin{eqnarray}\label{effedue}
\fl f(Q,\gamma,\chi,\widehat{\chi},m,\widehat{m})
=\frac{1}{2}nQ\widehat{\chi}
-\frac{1}{2}\gamma\chi-nm\widehat{m}\\+n\avg{\max_{s\geq
0}\l[-\frac{1}{2}\Delta \widehat{\chi}s^2+
\l(\widehat{m}+ t\sqrt{\gamma}\r)s\sqrt{\Delta}\r]}_{t,\Delta}
\nonumber\\ +\avg{\max_{x\geq 0}
\l[u(x)-\frac{1}{2\chi}\l(x-y+t\sqrt{nQ}+\eta nm\r)^2\r]}_{t,y}
\end{eqnarray}
where we used the same notation for averages and variables as in
Sec. 3.  The effective problems for firms and consumers are easily
identifiable and can be analyzed along the lines followed in
Sec. 3.1. We jump straight to the saddle-point equations, which, using
the same notation as in Sec. 3, now read
\begin{eqnarray}
m=\avg{s^*\sqrt{\Delta}}_{t,\Delta}\\
Q=\avg{(s^*)^2\Delta}_{t,\Delta}\\
\chi=\frac{n}{\sqrt{\gamma}}\avg{ts^*\sqrt{\Delta}}_{t,\Delta}\\
\gamma=\avg{u'(x^*)^2}_{t,y}\\
\widehat{m}=-\eta\avg{u'(x^*)}_{t,y}\\
\widehat{\chi}=\frac{1}{\sqrt{nQ}}\avg{t u'(x^*)}_{t,y}
\end{eqnarray}
It is clear that $m$ represents the average scale of operations while
$\widehat{m}$ is connected to the average (relative) price and
$\gamma$ is related to price fluctuations. The above equations can
again be solved numerically upon varying $n$. Adopting the same
choices as in (\ref{parama}), one obtains the behaviour shown in
Fig.~\ref{mQphi}.
Strikingly different macroscopic properties emerge from this model. In
particular, the average scale of operations diverges at a critical
point that for $\eta=1$ is roughly $n_c\simeq 2.1$. It is clear that
such an economy is completely unrealistic. When $m$ diverges, the
average relative price drops to zero, so that arbitrarily large
consumptions can be achieved (the reader can verify that relative
price fluctuations are nevertheless finite above $n_c$). We will show
in the next section that this behaviour is due to the fact that for
$n>n_c$ technologies can be combined in such a way that $\sum_i
s_i\xi_i^\mu>0$ for all $\mu$, i.e. that it is possible to produce
every good without consuming.

\subsection{The critical point}

We will now proceed by calculating the typical volume of configuration
space where technologies can be combined to yield $\sum_i
s_i\xi_i^\mu>0$ for all $\mu$. For a fixed disorder realization
$\boldsymbol{\xi}$, this volume is given by
\begin{equation}
V(\boldsymbol{\xi})=\int \prod_{\mu=1}^P
\theta\l(\sum_{i=1}^N s_i \xi_i^\mu\r) d\boldsymbol{s}
\end{equation}
For $n<n_c$, it is to be expected that $V(\boldsymbol{\xi})$
is non-zero only with an exponentially small (in $N$) probability,
whereas the probability of it being finite becomes of order $1$ at
$n_c$. The typical volume is obtained from the quenched average of
$V(\boldsymbol{\xi})$:
\begin{equation}
V_{{\rm typ}}=\exp \avg{\log
V(\boldsymbol{\xi})}_{\boldsymbol{\xi}}
\end{equation}
We will focus in particular on the quantity
\begin{equation}
S=\lim_{N\to\infty}\frac{1}{N}\log V_{{\rm typ}}
\end{equation}
(analogous to a zero-temperature entropy in statistical physics). From
the above discussion, we expect $S$ to be negative for $n<n_c$
and positive for $n>n_c$, so that the critical point $n_c$ can be
determined from the condition $S=0$.

In principle, $n_c$ depends on $\eta$ and $g(\Delta)$.  Notice,
however, that because of the statistics of the $\xi_i^\mu$'s, a
re-scaling $s_i\to s_i\sqrt{\Delta_i}$ makes the integral independent
of $\Delta_i$'s, apart from a trivial factor
$\prod_i\case{1}{\sqrt{\Delta_i}}$. Hence we shall set $\Delta_i=1$
for all $i$ without any loss of generality.

In order to ensure that integrals are well-defined, we introduce a
spherical constraint of the form $\sum_i s_i^2=N$ and concentrate on
\begin{equation}
V(\boldsymbol{\xi})=\int
e^{\ii\sum_{\mu}\widehat{x}^\mu(x^\mu-\sum_i s_i
\xi_i^\mu)-\frac{1}{2}\widehat{y}(\sum_i s_i^2-N)}
d\boldsymbol{s}d\boldsymbol{x}d\boldsymbol{\widehat{x}}
d\boldsymbol{\widehat{y}}
\end{equation}
As before, the evaluation of the $\boldsymbol{\xi}$-average requires
replication techniques. The calculation is standard and we don't
detail it here. In the limit $N\to\infty$ it ultimately turns out that
$S$ is given by the saddle-point value of
\begin{eqnarray}
\fl h(q,m,\widehat{m},r,y)=\frac{1}{2}qr-m\widehat{m}+\frac{1}{2}y^2
-\frac{1}{n}\log 2 \\+\frac{1}{2}\log\frac{\pi}{2(y+r)}+\frac{1}{n}
\avg{\log\l[1-\erf\l(\frac{\eta
      nm+t\sqrt{nq}}{\sqrt{2n(1-q)}}\r)\r]}_t\\
+\frac{\widehat{m}^2+r}{ 2(y+r)}+\avg{\log\l[1+
    \erf\l(\frac{\widehat{m}+t\sqrt{r}}{\sqrt{2(y+r)}}\r)\r]}_t
\end{eqnarray}
$n_c$ can be found by solving simultaneously the saddle-point
equations $\case{\partial h}{\partial q}=\cdots=\case{\partial
h}{\partial y}=0$ with $n=n_c$ together with the condition that $h$
vanishes for $n=n_c$.  This leads to the critical line $n_c(\eta)$
shown in Fig.~\ref{pd} (dashed line) for both positive and negative
values of $\eta$. The numerical value obtained for $\eta=1$, namely
$n_c=2.11524\ldots$, is in convincing agreement with the results of
the previous section. We remark that $n_c\to 1$ as $\eta\to
0$. Indeed, when $n\geq 1$ there are enough random vectors
$\boldsymbol{\xi}_i$ to span the entire $P$-dimensional space. On the
other hand, $n_c\sim \eta^2$ for $\eta\gg 1$.

\section{Conclusions}

The model we analyzed here is clearly unrealistic in many
respects. However it shows that tools of statistical physics of
disordered systems allow us to characterize the collective statistical
properties of a complex economy, beyond what na\"\i ve intuition and
simple probabilistic arguments would suggest. It is reasonably easy to
anticipate a change of behavior for $n\approx 2$ on the basis of
simple geometric arguments \cite{long}. It is hard however to
generalize these insights to derive, for example, the full picture of
the phase transition for $\eta\to 0$ we have described, or the
behavior of the economy in the two phases.

It would be desirable to extend this approach to the case with many
consumers or to analyze the effects induced by imperfect competition
(oligopolies, see e.g. \cite{MasColell}). A further extension to time
dependent economies may provide interesting insights into economic
growth theory. In particular, one can think about technological
improvement as a dynamics of the $\xi_i^\mu$'s occurring on time
scales much longer than those over which agents achieve
equilibrium. This would require a theory just moderately more
sophisticated than the one presented here.

\ack We are indebted to D Fiaschi, KG Maler, F Vega-Redondo and MA
Virasoro for important discussions. This work was partially supported
by the EU EXYSTENCE network and by the EU Human Potential Programme
under contract HPRN-CT-2002-00319, STIPCO.

\end{document}